\documentclass[12pt]{article}
\usepackage{epsfig,amsfonts,amssymb}
\usepackage{hyperref}
\usepackage{cite}
\input epsf.sty
\usepackage{cite}

\def\ZZZ{{\hbox{ Z\kern-1.6mm Z}}}
\def\RRR{{\hbox{ R\kern-2.4mm R}}}
\def\CCC{{\hbox{ C\kern-2.0mm C}}}
\def\zzz{{\hbox{z\kern-1mm z}}}

\newcommand{\qeq}{{\hbox{=\kern-2.3mm ? \kern.5mm }}}
\renewcommand{\qeq}{=}

\newcommand{\vp}{\varphi}

\newcommand{\GG}{{\cal G}}

\newcommand{\HH}{{\cal H}}
\newcommand{\MM}{{\cal M}}

\newcommand{\PP}{{\cal P}}

\newcommand{\XX}{{\cal X}}

\newcommand{\wt}{\widetilde}
\newcommand{\wh}{\widehat}

\newcommand{\RR}{{\cal R}}

\newcommand{\be}{\begin{equation}}
\newcommand{\ee}{\end{equation}}
\newcommand{\ben}{\begin{eqnarray}\displaystyle}
\newcommand{\een}{\end{eqnarray}}

\newcommand{\refb}[1]{(\ref{#1})}
\newcommand{\p}{\partial}
\newcommand{\sectiono}[1]{\section{#1}\setcounter{equation}{0}}

\def\one{{\hbox{ 1\kern-.8mm l}}}
\def\zero{{\hbox{ 0\kern-1.5mm 0}}}

\newcommand{\bea}[1]{\begin{eqnarray}\label{#1} }
\newcommand{\eea}{\end{eqnarray}}

\newcommand{\eqref}{\refb}

\begin{document}

\begin{flushright}
\end{flushright}

\vskip 12pt

\baselineskip 24pt

\begin{center}
{\Large \bf  BV Master Action for Heterotic and Type II String Field Theories}

\end{center}

\vskip .6cm
\medskip

\vspace*{4.0ex}

\baselineskip=18pt

\centerline{\large \rm Ashoke Sen}

\vspace*{4.0ex}

\centerline{\large \it Harish-Chandra Research Institute}
\centerline{\large \it  Chhatnag Road, Jhusi,
Allahabad 211019, India}

\vspace*{1.0ex}
\centerline{\small E-mail:  sen@mri.ernet.in}

\vspace*{5.0ex}

\centerline{\bf Abstract} \bigskip

We construct the quantum BV master action for heterotic and type II
string field theories.

\vfill \eject

\baselineskip=18pt

\tableofcontents

\sectiono{Introduction} 

Conventional formulation of superstring theory is based on an on-shell formulation in which
the S-matrix of on-shell external states are expressed as correlation functions of conformally
invariant vertex operators on a Riemann surface integrated over the moduli space of the
Riemann surface. However this approach is not suitable for addressing many issues even
within perturbation theory -- this includes the problem of mass renormalization and vacuum shift.
In the conventional approach, 
both these problems show up as infrared divergences associated with separating
type degenerations of the Riemann surfaces, but there is no suggested cure for this in 
perturbation theory\cite{1209.5461,1304.2832}. 
In recent papers \cite{1311.1257,1401.7014,1404.6254,1408.0571,1411.7478,1501.00988,1508.02481} 
we proposed a possible resolution of these problems based on one particle
irreducible (1PI) effective 
action in which the world-sheet theory is used to first construct a gauge invariant 1PI 
effective string field theory, and then we use this 1PI action to address the problem of finding the
vacuum and the 
renormalized masses following the usual route of quantum field theory. 
Since the 1PI action itself does not include contributions from separating type
degenerations of the Riemann surface, it does not suffer from any infrared divergences
associated with these degenerations. This makes this approach well-suited for addressing
the origin and resolution of these divergences in the S-matrix.

However the 1PI effective action does receive contribution from regions of the moduli space
associated with non-separating type degenerations. This makes it difficult to address issues related
to such divergences using the 1PI action since these divergences are hidden in
the building blocks of the theory -- the 1PI amplitudes.
For this reason it is useful to 
look for a field theory of strings in which the amplitudes are built from the Feynman diagrams 
of this string field theory.  In this formalism the elementary vertices will be free from infrared divergences
associated with both separating and non-separating type degenerations, and all infrared divergences
will appear when we build Feynman diagrams using these vertices.
This will make all the infrared divergences
manifest in perturbation theory, making it easier to use conventional field theory tools to analyze
the effect of these infrared divergences.
Needless to say, such a formulation also has the potential of
opening the path to studying non-perturbative aspects
of string theory.

For bosonic string theory, there has been successful construction of a field theory of open
strings as well as closed strings based on the Batalin-Vilkovisky (BV) 
formalism\cite{wittensft,thorn,bochicchio,saadi,kugo,9206084,9301097}.
There have been many attempts in the past to formulate a field theory of closed superstrings
/ heterotic strings, but  for various reasons, none has been completely successful at the quantum
level (see \cite{wittenssft,9503099,0109100,0406212,0409018,1201.1761,
1305.3893,1312.2948,1312.7197,1403.0940,
1407.8485,1412.5281,1505.01659,1505.02069,1506.05774,1506.06657,1508.00366} for a partial list of 
references). 
In this paper
we generalize the approach of \cite{1411.7478,1501.00988} 
for the construction of closed bosonic string field theory to construct
heterotic and type II string field theories.

The main difficulty in constructing a field theory for heterotic and type II strings has been in the Ramond
sector since there is no natural way to write down a kinetic term involving Ramond sector 
fields.\footnote{A recent proposal for dealing with this problem in classical open superstring field
theory can be
found in \cite{1508.00366}.} 
In the context 
of 1PI effective action, this
problem was recently addressed in \cite{1501.00988} using additional fields in the 
Ramond sector and then imposing a 
constraint on the external states that removes the extra states associated with these additional fields.
The combination on which we impose the constraint satisfies free field equations of motion, and hence
once we set them to zero, they are not produced by interactions. This makes the whole procedure
consistent, leading to a set of off-shell amplitudes satisfying the desired Ward identities. This was then
used to address the problem of computing renormalized masses, and also
computing amplitudes around the shifted
vacuum in cases where the perturbative vacuum is destabilized by quantum corrections.

The main observation we make in this paper is that the same trick can be used to construct a BV master
action for heterotic and type II string field theory. At the level of the classical theory itself, 
we introduce an additional set of fields. This doubles the number of degrees of freedom.  The
resulting gauge invariant theory has the sector that describes correctly the spectrum and interaction of
string theory  known from the first quantized approach, but there is an additional sector containing 
free fields. This theory can be quantized using BV formalism following the same procedure as in the case
of closed bosonic string field theory, but the quantum theory will also have the additional sector containing
free fields. At the end we are free to set the free fields to zero since they are never produced in any 
interactions (i.e.\ in the scattering involving external states in the interacting sector, the additional fields 
will never be produced as intermediate states).

We shall not try to make the paper self-contained. Instead we shall assume that
the reader is familiar with the construction of the BV master action in closed
string field theory\cite{9206084}. Some familiarity with the construction of  the 1PI
action in superstring field theory is also desirable, although
we give a brief
review of some of the results of \cite{1411.7478,1501.00988} in \S\ref{sreview}. In \S\ref{s3} 
we describe the
construction of the action satisfying classical master equation. In \S\ref{s4} we describe the
construction of the full quantum master action and its gauge fixing.

\sectiono{Review} \label{sreview}

Since our construction will follow closely the conventions used in \cite{1411.7478,1501.00988}, 
we shall not give a detailed
review of the background material, but only describe a few ingredients that will be used in our
analysis. A more detailed review can be found {\it e.g.} in \cite{1508.02481}.
This section will contain three parts. In \S\ref{s2.1} we review some of the 
details of the
superconformal field theory (SCFT) describing the world-sheet theory of the matter and ghost 
system\cite{FMS}. 
In \S\ref{sbrac} we review the construction of certain
multilinear functions of states of the SCFT and how we use them to 
construct the 1PI effective action.
In \S\ref{s2.2} we describe the construction of classical (tree level) 
string field theory from the 1PI action.  This classical action will then be used in 
\S\ref{s3} for the construction of the classical master action, which will then be 
generalized to quantum master action in \S\ref{s4}.
  
\subsection{The world-sheet theory}  \label{s2.1}

We denote by $\HH$ the full Hilbert space of matter ghost SCFT carrying arbitrary picture and
ghost numbers, and by $\HH_T$ a subspace of $\HH$ satisfying the constraints
\be 
b_0^- |s\rangle =0, \quad L_0^- |s\rangle =0, \quad \hbox{for $|s\rangle \in \HH_T$}
\, ,
\ee
where
\be
b_0^\pm = b_0\pm \bar b_0, \quad L_0^\pm = L_0\pm \bar L_0, \quad c_0^\pm ={1\over 2}
(c_0\pm \bar c_0)\, .
\ee
We denote by $\XX$ the picture changing operator (PCO) -- for type II string theories we also have
its anti-holomorphic counterpart $\bar\XX$. $\XX_0$ and $\bar\XX_0$ are their zero modes\cite{9711087,1312.2948,1403.0940}:\footnote{In \cite{1508.00366} 
a different operator with properties similar to $\XX_0$
was used.} 
\be
\XX_0 =\ointop {dz\over z} \XX(z), \quad \bar\XX_0 =\ointop {d\bar z\over \bar z} \bar\XX(\bar z) \quad
\hbox{(in type II)}\, .
\ee
In heterotic theory we divide $\HH_T$ into Neveu-Schwarz (NS)
sector $\HH_{NS}$ and Ramond (R) sector $\HH_R$. In the type II theory the corresponding
division is $\HH_{NSNS}$, $\HH_{NSR}$, $\HH_{RNS}$ and $\HH_{RR}$. The operator 
$\GG$ in these theories is defined as
\be \label{edefgg}
\GG|s\rangle =\begin{cases} {|s\rangle \quad \hbox{if $|s\rangle\in \HH_{NS}$}\cr
\XX_0\, |s\rangle \quad \hbox{if $|s\rangle\in \HH_R$}}
\end{cases}\, ,
\ee
in heterotic string theory, and
\be \label{edefggii}
\GG|s\rangle =\begin{cases} {|s\rangle \quad \hbox{if $|s\rangle\in \HH_{NSNS}$}\cr
\XX_0\, |s\rangle \quad \hbox{if $|s\rangle\in \HH_{NSR}$}\cr 
\bar\XX_0\, |s\rangle \quad \hbox{if $|s\rangle\in \HH_{RNS}$}\cr 
\XX_0\bar\XX_0\, |s\rangle \quad \hbox{if $|s\rangle\in \HH_{RR}$}\cr 
}\, ,
\end{cases}
\ee
in type II theory. It satisfies
\be \label{eqbgg}
[Q_B, \GG]=0, \quad [b_0^\pm, \GG] = 0\, .
\ee

The basis states in $\HH_{NS}$ are taken to be grassmann even for 
even ghost number
and grassmann odd for odd ghost number. In $\HH_R$ the situation is opposite.
In type II string theory the basis states are grassmann even for 
even ghost number
and grassmann odd for odd ghost number in $\HH_{NSNS}$ and $\HH_{RR}$.
In $\HH_{RNS}$ and $\HH_{NSR}$ the situation is opposite.

In heterotic string theory, 
we denote by $\wh\HH_T$ the subspace of $\HH_T$ 
containing states of picture numbers 
$-1$  and $-1/2$ in the NS and R sectors respectively. 
$\wt\HH_T$ will denote the subspace of states of picture numbers
$-1$ and $-3/2$ in the NS and R sectors. In type II theories $\wh \HH_T$ will contain states of picture
numbers $(-1,-1)$, $(-1,-1/2)$, $(-1/2, -1)$ and $(-1/2, -1/2)$ in the NSNS, NSR, RNS and RR
sectors while $\wt\HH_T$ will contain states of picture
numbers $(-1,-1)$, $(-1,-3/2)$, $(-3/2, -1)$ and $(-3/2, -3/2)$ in the NSNS, NSR, RNS and RR
sectors.

\subsection{The 1PI action} \label{sbrac}

Generalizing the construction of 
\cite{9206084} for closed bosonic string theory, in 
\cite{1408.0571,1411.7478,1501.00988} we introduced, for the heterotic string, 
the space $\wt\PP_{g,m,n}$ 
whose base was the moduli space 
$\MM_{g,m,n}$ of genus $g$ Riemann surface with $m$ NS and $n$ R
punctures and whose fiber  contains information on local
coordinates up to phases and also  the locations
of $(2g-2+m + n/2)$ PCO's. 
(The generalization to type II strings is straightforward.)
Furthermore for $m$ external NS sector states and
$n=N-m$ external R-sector states in $\wh\HH_T$, collectively called
$|A_1\rangle,\cdots |A_N\rangle$, we introduced on $\wt\PP_{g,m,n}$ a $p$-form
$\Omega^{(g,m,n)}_p(|A_1\rangle,\cdots |A_N\rangle)$ 
for all integer $p\ge 0$
satisfying certain desired properties. Finally for each $g,m,n$ we introduced
a specific subspace of $\MM_{g,m,n}$ and (generalized) section\footnote{Generalized sections include weighted 
average of sections. Furthermore they may contain `vertical segments'
in which the PCO locations may jump 
discontinuously across codimension 1 subspaces in the interior of 
$\RR_{g,m,n}$\cite{1408.0571,1504.00609}. \label{f1}} 
$\RR_{g,m,n}$ of $\wt\PP_{g,m,n}$ on 
these subspaces satisfying the conditions
\ben  \label{eboundary}
\p \RR_{g,m,n} &=& -{1\over 2} \sum_{g_1,g_2\atop g_1+g_2=g} 
\sum_{m_1,m_2\atop m_1+m_2 = m+2}
\sum_{n_1,n_2\atop n_1+n_2 = n}
{\bf S}[\{\RR_{g_1,m_1,n_1} , \RR_{g_2,m_2,n_2}\}] \nonumber \\ &&
-{1\over 2} \sum_{g_1,g_2\atop g_1+g_2=g} 
\sum_{m_1,m_2\atop m_1+m_2 = m}
\sum_{n_1,n_2\atop n_1+n_2 = n+2}
{\bf S}[\{\RR_{g_1,m_1,n_1} ; \RR_{g_2,m_2,n_2}\}]
\, .
\een
Here $\p \RR_{g,m,n}$ denotes the boundary of $\RR_{g,m,n}$
and ${\bf S}$ denotes the
operation of summing over inequivalent permutations of external NS-sector punctures and
also external R-sector punctures.
$\{\RR_{g_1,m_1,n_1} , \RR_{g_2,m_2,n_2}\}$ denotes the  
subspace of $\wt\PP_{g_1+g_2, m_1+m_2-2, n_1+n_2}$ obtained 
by gluing the Riemann surfaces in $\RR_{g_1.m_1.n_1}$ and
$\RR_{g_2,m_2, n_2}$ at one NS puncture from each via the 
{\it special} plumbing
fixture relation\footnote{These correspond to $s=0$ boundaries of the 
general plumbing fixture relations given in \refb{eplumbing}.}
\be \label{eplumbspecial}
z \, w = e^{i\theta},\quad 0\le\theta \le 2\pi\, ,
\ee
where $z$ and $w$ denote local coordinates around the punctures that are
being glued.
Similarly $\{\RR_{g_1,m_1,n_1} ; \RR_{g_2,m_2,n_2}\}$ denotes the  
subspace of $\wt\PP_{g_1+g_2, m_1+m_2, n_1+n_2-2}$ obtained 
by gluing the Riemann surfaces in $\RR_{g_1,m_1,n_1}$ and
$\RR_{g_2,m_2, n_2}$ at one R puncture from each via the same
special plumbing
fixture relation \refb{eplumbspecial}. 
There is one additional subtlety in the definition of $\{~;~\}$. The total
number of PCO's on the two 
Riemann surfaces corresponding to a point
in $\RR_{g_1, m_1, n_1}$ and a point in $\RR_{g_2,m_2,n_2}$ is 
$2(g_1+g_2)-4 + (m_1+m_2) + (n_1+n_2)/2$. Using the constraints
given in the second term in \refb{eboundary}, this can be written as
$(2g-2) + m + n/2 -1$, which is one less than the required number of
PCO's on a Riemann surface associated with a point in $\wt\PP_{g,m,n}$.
Therefore in defining $\{~;~\}$ we need to prescribe the location of the 
additional PCO. A consistent prescription that we shall adopt is to insert
a factor of $\XX_0$ around one of the two punctures which are being 
glued. Which of the two punctures we choose is irrelevant since
$\ointop dz\, z^{-1}\, \XX(z) =\ointop dw\, w^{-1}\, \XX(w)$ when $z$
and $w$ are related as in \refb{eplumbspecial}. In fact in both heterotic
and type II string theories, a universal prescription for plumbing fixture
rules in all sectors 
will be to insert the operator $\GG$ defined in \refb{edefgg}, 
\refb{edefggii} at one
of the two punctures which are being glued. 

$\RR_{g,m,n}$'s can be called
`1PI subspaces' of $\wt\PP_{g,m,n}$ since, as we shall see, they can be
used to define 1PI amplitudes.
Operationally the regions $\RR_{g,m,n}$ are constructed as follows.
For $(g=0, m+n=3)$ and $(g=1,m+n=1)$ we choose $\RR_{g,m,n}$
so that its projection 
to $\MM_{g,m,n}$
is the whole moduli space $\MM_{g,m,n}$ and the choice of the section
encoding choice of local coordinates and PCO locations are arbitrary
subject to symmetry restrictions -- permutations of punctures for $(g=0,m+n=3)$
and modular invariance for $(g=1,m+n=1)$.  For $(g=1,m+n=1)$ the section must also
avoid spurious poles\cite{Verlinde:1987sd,lechtenfeld,morozov}.
Achieving these may involve making use
of generalized sections in the sense described in footnote \ref{f1}.
Given these choices we now
glue the Riemann surfaces corresponding to points in
these $\RR_{g,m,n}$'s via the plumbing 
fixture relations 
\be \label{eplumbing}
z \, w = e^{-s+i\theta},\quad 0\le s <\infty, \quad 0\le\theta \le 2\pi\, .
\ee
While carrying
out the plumbing fixture we always choose a pair of punctures
on {\it two different Riemann surfaces} -- we never use a pair of punctures
on the same Riemann surface. In the first stage
these generate subspaces of $\wt\PP_{g,m,n}$ for $(g=0, m+n=4)$ and
$(g=1, m+n=2)$ -- we ignore the $(g=2,m+n=0)$ sector since the associated
Riemann surface has no punctures where the vertex operators can be inserted.
Typically the projection of these subspaces to $\MM_{g,m,n}$ 
do not cover the whole of
$\MM_{g,m,n}$ for these values of $(g,m,n)$. 
We choose the $\RR_{g,m,n}$ for $(g=0,m+n=4)$ and
$(g=1, m+n=2)$ so as to `fill these gaps'.  Only the boundary of
$\RR_{g,m,n}$ is fixed from this consideration;
how we fill the gap is arbitrary, except that we choose them in a manner
consistent with the various symmetries {\it e.g.} exchange of the NS punctures
and exchange of the R punctures and also avoiding spurious 
poles. 
The requirement that the boundaries
of the new regions $\RR_{g,m,n}$ match the $s=0$
boundaries of the regions
of $\wt\PP_{g,m,n}$ obtained by plumbing fixture of Riemann surfaces
associated with $\RR_{g,m,n}$ with $(g=0,m+n=3)$ and $(g=1,m+n=1)$
leads to the conditions  \refb{eboundary}. We now continue this process, 
generating new subspaces of $\wt\PP_{g,m,n}$ by plumbing fixture of the
subspaces $\RR_{g',m',n'}$ that have already been determined.
We allow the Riemann surfaces associated with 
these subspaces to be glued 
multiple number of times, but ensuring that at no stage
we glue two punctures situated on the same Riemann surface. We then define
new $\RR_{g,m,n}$'s by filling the gap left-over from this construction.
Continuing this process we 
construct all the $\RR_{g,m,n}$'s. 

Once $\RR_{g,m,n}$'s have been constructed this way, we define
a multilinear function 
$\{A_1\cdots A_N\}$  of $|A_1\rangle,\cdots |A_N\rangle\in \wh\HH_T$
via the relation
\be \label{edefcurly}
\{A_1\cdots A_{m+n}\}= \sum_{g=0}^\infty (g_s)^{2g} 
\int_{\RR_{g,m,n}} \Omega^{(g,m,n)}_{6g-6+2m+2n}(|A_1\rangle,
\cdots |A_{m+n}
\rangle)\, .
\ee
Physically these represent 1PI amplitudes with external states
$|A_1\rangle,\cdots |A_N\rangle$.
We also introduced
another multilinear function 
$[A_2\cdots A_N]$ of $|A_2\rangle,\cdots |A_N\rangle\in \wh\HH_T$
taking values in $\wt\HH_T$ defined via
\be \label{edefsquare}
\langle A_1| c_0^- |[A_2\cdots A_N]\rangle = \{A_1\cdots A_N\}
\ee
for all $|A_1\rangle\in\wh\HH_T$. Here $\langle A|B\rangle$ denotes the BPZ inner
product between two states $|A\rangle$ and $|B\rangle$ in the full Hilbert space $\HH$.
These functions  satisfy the identities
\be \label{ecurlysym}
\{A_1 A_2\cdots A_{i-1}A_{i+1} A_iA_{i+2} \cdots A_N\}
=(-1)^{\gamma_i \gamma_{i+1}} \{A_1A_2\cdots A_N\}\, ,
\ee
\be \label{esymmetry}
[A_1\cdots A_{i-1}A_{i+1} A_iA_{i+2} \cdots A_N]
=(-1)^{\gamma_i \gamma_{i+1}} [A_1\cdots A_N] \, ,
\ee
where $\gamma_i$ is the grassmannality of $|A_i\rangle$.
They also satisfy
\ben \label{eimpid}
&&  \sum_{i=1}^N (-1)^{\gamma_1+\cdots \gamma_{i-1}}\{A_1\cdots A_{i-1} (Q_B A_i)
A_{i+1} \cdots A_N\} \nonumber \\
&=& -  
{1\over 2} \sum_{\ell,k\ge 0\atop \ell+k=N} \sum_{\{i_a;a=1,\cdots \ell\}, \{j_b;b=1,\cdots k\}\atop
\{i_a\}\cup \{j_b\} = \{1,\cdots N\}
}\sigma(\{i_a\}, \{j_b\})
\{A_{i_1} \cdots A_{i_\ell}\GG[A_{j_1} \cdots A_{j_k}]\}
\, 
\een
and
\ben \label{emain}
&& Q_B[A_1\cdots A_N] + \sum_{i=1}^N (-1)^{\gamma_1+\cdots 
\gamma_{i-1}}[A_1\cdots A_{i-1} (Q_B A_i)
A_{i+1} \cdots A_N] \nonumber \\
&=& -  \sum_{\ell,k\ge 0\atop \ell+k=N} 
\sum_{\{i_a;a=1,\cdots \ell\}, \{j_b;b=1,\cdots k\}\atop
\{i_a\}\cup \{j_b\} = \{1,\cdots N\}
}\sigma(\{i_a\}, \{j_b\})\, 
[A_{i_1} \cdots A_{i_\ell} \GG\, [A_{j_1} \cdots A_{j_k}]]
\een
where $\sigma(\{i_a\}, \{j_b\})$ is the sign that one picks up while rearranging
$b_0^-,A_1,\cdots A_N$ to\break \noindent
$A_{i_1},\cdots A_{i_\ell}, b_0^-, A_{j_1},\cdots A_{j_k}$.
Finally we also have a relation
\be \label{eneweq2a}
\{A_1\cdots A_k \GG[\wt A_1\cdots \wt A_\ell]\}
= (-1)^{\gamma+\wt\gamma+\gamma\wt\gamma}
\{\wt A_1\cdots \wt A_\ell \GG[A_1\cdots A_k]\}\, ,
\ee
where $\gamma$ and $\wt\gamma$ are the total grassmannalities of
$A_1,\cdots A_k$ and $\wt A_1,\cdots \wt A_\ell$ respectively.

These ingredients can be used to construct the 1PI action of the theory as
follows\cite{1411.7478,1501.00988}. 
We take the string field to consist of two components $|\Psi\rangle$ and
$|\wt\Psi\rangle$. $|\Psi\rangle$ is taken to be an arbitrary element of ghost number 2
in $\wh\HH_T$ and $|\wt\Psi\rangle$ is taken to be an arbitrary element of ghost number 2
in $\wt\HH_T$. Both string fields are taken to be grassmann even. It follows from the
paragraph below \refb{eqbgg} that in the heterotic string
theory the expansion coefficients are grassmann even for $\HH_{NS}$ and
grassmann odd for $\HH_R$, while in type II string theory the expansion coefficients are grassmann 
even for $\HH_{NSNS}$ and $\HH_{RR}$ and
grassmann odd for $\HH_{NSR}$ and $\HH_{RNS}$.
The 1PI action has the form
\be \label{e1PIact}
S_{1PI}= g_s^{-2}\left[
-{1\over 2} \langle\wt\Psi |c_0^- Q_B \GG |\wt\Psi\rangle 
+ \langle\wt\Psi |c_0^- Q_B |\Psi\rangle + 
\sum_{n=1}^\infty {1\over n!} \{ \Psi^n\}
\right] \, ,
\ee
where  $g_s$ denotes string coupling and $\{\Psi^n\}$ means
$\{ \Psi\cdots \Psi\}$ with $n$ insertions of $|\Psi\rangle$. 
It is easy to see that the action 
\refb{e1PIact} is invariant under the infinitesimal gauge transformation 
\be \label{egaugepsi}
|\delta\Psi\rangle = Q_B|\Lambda\rangle + \sum_{n=0}^\infty {1\over n!} 
\GG [\Psi^n \Lambda]\, , \quad
|\delta\wt\Psi\rangle = Q_B |\wt\Lambda\rangle + \sum_{n=0}^\infty {1\over n!} 
[\Psi^n \Lambda]\, ,
\ee
where $|\Lambda\rangle\in\wh\HH_T$, $|\wt\Lambda\rangle\in \wt\HH_{T}$, 
and both carry ghost number 1.

The 1PI action given in
\refb{e1PIact} is not unique but depends on the choice of 
$\RR_{g,m,n}$, i.e.\ choice of local coordinates 
at the punctures and PCO locations.  Different choices
lead to different definitions of $\{A_1\cdots A_N\}$. However the corresponding 
1PI effective string field
theories can be shown to be related by field redefinition, and hence
this ambiguity does not affect any of the physical quantities. While we shall not
make any specific 
assumption about the choice of local coordinates and PCO locations, we shall
assume that the local coordinates have been scaled by a sufficiently large 
number so that unit radius circle around the punctures in the local coordinates
correspond to physically small disks around the punctures\footnote{In string field
theory literature this is often described as adding long stubs to the external
lines of the vertex.}, and that the PCO's
are inserted outside these unit disks. This will ensure that in the 1PR amplitudes
obtained by gluing the 1PI amplitudes via \refb{eplumbing}, 
the PCO's do not collide. This also ensures that 
as long as the 1PI amplitudes $\{A_1\cdots A_N\}$
are free from spurious singularities, the 1PR amplitudes built from plumbing 
fixture of these 1PI amplitudes 
are also free from spurious singularities.

\subsection{Classical action} \label{s2.2}

For the construction of the classical action we can 
restrict our attention to only the genus zero
contribution to the functions $\{A_1\cdots A_N\}$ and $[A_2\cdots A_N]$, 
which we shall denote by 
$\{A_1\cdots A_N\}_0$ 
and $[A_2\cdots A_N]_0$ respectively. These
functions vanish for $N\le 2$.
The classical action of the theory can now be written down from the 1PI effective
action \refb{e1PIact} using the fact that at tree level there is no difference between
the classical action and the 1PI action. Therefore it takes the form
\be \label{eactpsi}
S_{\rm cl} = g_s^{-2}\left[
-{1\over 2} \langle\wt\Psi |c_0^- Q_B \GG |\wt\Psi\rangle 
+ \langle\wt\Psi |c_0^- Q_B |\Psi\rangle + 
\sum_{n=3}^\infty {1\over n!} \{ \Psi^n\}_0
\right] \, ,
\ee
with the gauge transformation taking the form
\be \label{egaugepsiclass}
|\delta\Psi\rangle = Q_B|\Lambda\rangle + \sum_{n=1}^\infty {1\over n!} 
\GG [\Psi^n \Lambda]_0\, , \quad
|\delta\wt\Psi\rangle = Q_B |\wt\Lambda\rangle + \sum_{n=1}^\infty {1\over n!} 
[\Psi^n \Lambda]_0\, .
\ee
The equations of motion derived from \refb{eactpsi} can be written as
\be \label{e01psi}
Q_B (|\Psi\rangle - \GG|\wt\Psi\rangle) = 0\, ,
\ee
\be \label{e02psi}
Q_B |\wt\Psi\rangle + \sum_{n=3}^\infty {1\over (n-1)!} [\Psi^{n-1}]_0= 0\, .
\ee

A priori this theory has too many degrees of freedom. For example at the
linearized level, the gauge inequivalent solutions to \refb{e01psi}
and \refb{e02psi} are given by the elements of BRST cohomology in the 
ghost number 2 sectors of $\wh\HH_T$ and $\wt\HH_T$. This will double the
number of physical states.\footnote{The doubling trick for dealing with Ramond
sector in Berkovits version of open string field theory has been explored
previously in \cite{0412215}.  The relationship between our approach and 
the approach of
\cite{0412215} is not completely clear. In particular one of the key features
of our approach is that the field $\wt\Psi$ enters the action only in 
quadratic terms. This features seems to be absent in \cite{0412215}.
} 
To circumvent this difficulty we observe that 
given any solution to the equations of motion \refb{e01psi},
\refb{e02psi}, we can generate new solutions by adding 
to $|\wt\Psi\rangle$ arbitrary BRST invariant states
keeping $|\Psi\rangle$ fixed. 
This suggests the following two step process for solving the equations of
motion. First by adding $\GG$ operated on the second equation to the first
equation we write the independent equations as
\be \label{e03psi}
Q_B |\Psi\rangle + \sum_{n=3}^\infty {1\over (n-1)!} \GG[\Psi^{n-1}]_0= 0\, ,
\ee
and
\be \label{e04psi}
Q_B |\wt\Psi\rangle + \sum_{n=3}^\infty {1\over (n-1)!} [\Psi^{n-1}]_0= 0\, .
\ee
In the first step we find general solutions of \refb{e03psi} without any
reference to \refb{e04psi}, and then, for each of these solutions, pick a 
particular $|\wt\Psi\rangle$ that solves \refb{e04psi}.\footnote{One might wonder
whether given a solution to \refb{e03psi}, one can always find a solution to
\refb{e04psi}. One class of solutions to these equations may be obtained 
by starting with a seed solution to the linearized equations of motion carrying
some generic momentum,
and then correcting it iteratively using the general procedure 
described {\it e.g.} in \cite{1411.7478,1501.00988,1508.02481}. 
In this case one can
relate
possible obstruction to finding iterative solutions to these equations to the
question of whether or not the non-linear terms in the equations of motion are 
BRST trivial. Using
the isomorphism between BRST cohomologies in different picture number sector
for generic momenta given in \cite{9711087}, one can then show that if the non-linear
terms in \refb{e03psi} are BRST trivial, then the non-linear terms in \refb{e04psi} are
also BRST trivial. Therefore
given a solution
to \refb{e03psi} one can find a solution to \refb{e04psi}.  
This leaves open the possibility that there may be `large' classical solutions
to \refb{e03psi} for which there is no solution to \refb{e04psi}. In such cases
we can simply discard these solutions without violating anything that we know in
perturbative string theory.} 
We could
implement this by imposing some specific condition like 
$|\Psi\rangle - \GG|\wt\Psi\rangle=0$, but this will not be necessary. In the second
step, for each of the solutions obtained at the first step, we add to $|\wt\Psi\rangle$
an arbitrary element of the BRST cohomology in the ghost number 2 sector
of $\wt\HH_T$. This generates the most general solution to the full set of
equations of motion. 
Since the deformation of the solution generated in the second step 
do not get modified by interactions,
and do not affect the solution generated in the first step,  upon quantization
they will represent free particles which do not scatter with each other
or with the particles
associated with the solutions to \refb{e03psi}. Thus this sector
decouples from the theory at tree level. This can also be seen from the
analysis of Feynman diagrams\cite{1411.7478,1501.00988,1508.02481}.
It follows from the analysis of \cite{1411.7478,1501.00988,1508.02481} --
restricted to tree level string theory --
that the interacting part of the theory describes
correctly the spectrum and S-matrix of string theory at tree level.

The gauge inequivalent
solutions to the {\it linearized} equations of motion at
the first step are characterized by the elements of the BRST cohomology
in the ghost number two sector of $\wh\HH_T$, whereas 
the gauge inequivalent
solutions to the {\it linearized} equations of motion at
the second step are characterized by the elements of the BRST cohomology
in the ghost number two sector of $\wt\HH_T$. 
This shows that the physical states in the interacting part of the theory
are in the BRST cohomology in $\wh\HH_T$ while the physical states which 
decouple are in the BRST cohomology in $\wt\HH_T$. 
The two are isomorphic at
non-zero momentum, but not at zero momentum\cite{9711087}.

We shall see in eq.\refb{emasterquantum}
that the interaction terms in the action in the full quantum theory
continue to be independent of $|\wt\Psi\rangle$. 
Hence the particles associated with the modes where we deform
$|\wt\Psi\rangle$ by adding a BRST invariant state keeping
$|\Psi\rangle$ fixed 
will never appear as intermediate states in an amplitude even in the full quantum
theory.
This will be demonstrated explicitly in \S\ref{s4.2} where we shall
derive the Feynman rules in the full quantum theory. 
In what follows we shall work with the full classical
action \refb{eactpsi} and its quantum
generalization \refb{emasterquantum}
at intermediate stages, and discuss the decoupling of the modes
of $|\wt\Psi\rangle$ only at the very end. 

\section{Classical master action} \label{s3}

We shall now construct the classical master action corresponding to the BV
quantization of the action \refb{eactpsi}. We follow the procedure described in
\cite{9206084} for closed bosonic string field theory. This is done in several steps.
\begin{enumerate}
\item First we relax the constraint on the ghost number and let
$|\Psi\rangle$ and $|\wt\Psi\rangle$ be arbitrary states in $\wh\HH_T$
and $\wt\HH_T$. The grassmannality of the coefficients are chosen such
that the string field is always even.
\item We divide $\wh \HH_T$ and $\wt\HH_T$ into two subsectors: $\wh\HH_+$
and $\wt\HH_+$ will contain states in $\wh\HH_T$ and $\wt\HH_T$ of
ghost numbers $\ge 3$, while  $\wh\HH_-$
and $\wt\HH_-$ will contain states in $\wh\HH_T$ and $\wt\HH_T$ of
ghost numbers $\le 2$. We introduce basis states $|\wh\vp^-_r\rangle$,
$|\wt\vp^-_r\rangle$, $|\wh\vp_+^r\rangle$ and $|\wt \vp_+^r\rangle$ of
$\wh\HH_-$,
$\wt\HH_-$, $\wh\HH_+$
and $\wt\HH_+$ satisfying orthonormality conditions
\be \label{einner}
\langle \wh\vp^-_r|c_0^- |\wt \vp_+^s \rangle = \delta_r{}^s=
\langle \wt \vp_+^s|c_0^- |\wh\vp^-_r\rangle, \quad 
\langle \wt\vp^-_r|c_0^- |\wh \vp_+^s \rangle = \delta_r{}^s=
\langle \wh \vp_+^s|c_0^- |\wt\vp^-_r\rangle\, ,
\ee
and expand the string fields $|\Psi\rangle$, $|\wt\Psi\rangle$ as
\ben \label{ephiexpan}
|\wt\Psi\rangle &=& \sum_r |\wt\vp^-_r\rangle \wt\psi^r  
+\sum_r (-1)^{g_r^*+1} |\wt\vp_+^r\rangle \psi_r^*\, , \nonumber \\
|\Psi\rangle -{1\over 2} 
\GG|\wt\Psi\rangle &=& \sum_r |\wh\vp^-_r\rangle \psi^r  
+ \sum_r (-1)^{\wt g_r^*+1} |\wh\vp_+^r\rangle \wt\psi_r^* 
\, .
\nonumber \\
\een
Here $g^*_r$, $g_r$, $\wt g^*_r$ and $\wt g_r$ label the grassmann
parities of $\psi^*_r$, $\psi^r$, $\wt \psi^*_r$ and $\wt\psi^r$
respectively. They in turn can be determined from the assignment of grassmann
parities to the basis states as described below \refb{eqbgg} and the fact that
$|\Psi\rangle$ and $|\wt\Psi\rangle$ are both even.  
\item We shall identify the variables 
$\{\psi^r, \wt\psi^r\}$ as `fields' and the
variables $\{\psi^*_r, \wt\psi^*_r\}$ as the conjugate `anti-fields' in the BV
quantization of the theory. It can be easily seen
that $\psi^r$ and $\psi^*_r$
carry opposite grassmann parities and $\wt \psi^r$ and $\wt\psi^*_r$
carry opposite grassmann parities. This is consistent with their identifications
as fields and conjugate anti-fields.
\item Given two functions $F$ and $G$ of all the fields and anti-fields,
we now define their anti-bracket in the standard way:
\be \label{eantib}
\{F, G\}= {\p_R F\over \p \psi^r} \, {\p_L G\over \p\psi^*_r}
+ 
{\p_R F\over \p \wt\psi^r} \, {\p_L G\over \p\wt\psi^*_r}
- {\p_R F\over \p \psi^*_r} \, {\p_L G\over \p\psi^r}
-
{\p_R F\over \p \wt\psi^*_r} \, {\p_L G\over \delta\wt\psi^r}\, ,
\ee
where the subscripts $R$ and $L$ of $\p$ denote left and right derivatives 
respectively.
\item The anti-bracket can be given the following interpretation in the
world-sheet SCFT. 
Given a function $F(|\Psi\rangle, |\wt\Psi\rangle)$ let us define 
$\langle F_R|$, $\langle \wt F_R|$, $|F_L\rangle$, $|\wt F_L\rangle$ 
such that under an infinitesimal variation of $|\Psi\rangle$, $|\wt\Psi\rangle$
we have
\be \label{edeffrfl}
\delta F = \langle F_R |c_0^- | \delta \wt \Psi\rangle 
+ \langle \wt F_R |c_0^- | \delta \Psi\rangle 
= \langle \delta \wt\Psi | c_0^- | F_L\rangle + \langle \delta
\Psi | c_0^- | \wt F_L\rangle\, .
\ee
Then using completeness of the basis states and using 
\refb{einner}-\refb{eantib} one can show that
the anti-bracket between two functions $F$ and $G$ is given by
\be \label{eantibracket}
\{ F, G\} = -\left(\langle F_R | c_0^- |\wt G_L\rangle + \langle \wt F_R
|c_0^- | G_L\rangle
+ \langle \wt F_R | c_0^- \GG | \wt G_L\rangle\right)
\, .
\ee
\item The classical BV master action of string field theory is now taken to
be of the same form as \refb{eactpsi} but with $|\Psi\rangle$, $|\wt\Psi\rangle$
containing states of all ghost numbers:
\be \label{emaster}
S = g_s^{-2}\left[
-{1\over 2} \langle\wt\Psi |c_0^- Q_B \GG |\wt\Psi\rangle 
+ \langle\wt\Psi |c_0^- Q_B |\Psi\rangle + 
\sum_{n=3}^\infty {1\over n!} \{ \Psi^n\}_0
\right] \, .
\ee
\end{enumerate}

We shall now check that this action satisfies the classical master equation.
Using \refb{emaster} and \refb{edeffrfl} we get
\ben
&& \langle S_R| =- \langle \Psi|Q_B + \langle \wt\Psi| Q_B\GG, \quad 
\langle \wt S_R| = -\langle \wt \Psi| Q_B + \sum_{n=3}^\infty {1\over (n-1)!}
\langle[\Psi^{n-1}]_0| \, , \nonumber \\
&& |S_L\rangle = Q_B|\Psi\rangle - Q_B \GG|\wt\Psi\rangle, \quad
|\wt S_L\rangle = Q_B |\wt\Psi\rangle  + \sum_{n=3}^\infty {1\over (n-1)!}
|[\Psi^{n-1}]_0\rangle\, .
\een
Therefore from \refb{eantibracket} we have
\ben \label{emastermani}
\{S, S\} &=& -\left(\langle S_R | c_0^- |\wt S_L\rangle + \langle \wt S_R
|c_0^- | S_L\rangle
+ \langle \wt S_R | c_0^- \GG | \wt S_L\rangle\right)
\nonumber \\
&=& - 2 \sum_{n=3}^\infty {1\over (n-1)!}
\langle \Psi | c_0^- Q_B [\Psi^{n-1}]_0 \rangle 
- \sum_{m=3}^\infty \sum_{n=3}^\infty {1\over (m-1)! (n-1)!}
\langle \GG[\Psi^{m-1}]_0|c_0^- 
|[\Psi^{n-1}]_0\rangle \nonumber \\
&=& - 2  \sum_{n=3}^\infty {1\over (n-1)!} \{ \Psi^{n-1} Q_B\Psi\}_0
- \sum_{m=3}^\infty \sum_{n=3}^\infty {1\over (m-1)! (n-1)!}
\{ \GG[\Psi^{m-1}]_0
\Psi^{n-1}\}_0 \nonumber \\ 
&=& 0\, ,
\een
where in the last step we have used \refb{eimpid}.  Note that the $|\wt\Psi\rangle$ dependent terms
cancel in going from the first to the second line itself, and 
this cancelation does not require any
details of the interaction terms except that they depend only on $|\Psi\rangle$.
The manipulations leading from second to the fourth line are identical to
what is done in closed bosonic string field theory\cite{9206084}, 
except for
insertion of factor of $\GG$ on $[\cdots]_0$.
Eq.\refb{emastermani} shows that the
action $S$ satisfies the classical master equation $\{S, S\}=0$. 

\sectiono{Quantum master action} \label{s4}

\newcommand{\cL} {\{\hskip -4pt\{}
\newcommand{\cR} {\}\hskip -4pt\}}
\newcommand{\sL} {[\hskip -1.5pt[}
\newcommand{\sR} {]\hskip -1.5pt]}
\newcommand{\oR}{{\overline{\RR}}}


Given the construction of the classical master action and the definitions
of fields and anti-fields given in \S\ref{s3}, the construction of the
quantum master action can be given using the same steps as in
\cite{9206084}, with the necessary modifications for superstrings read out from
the results of \cite{1411.7478,1501.00988}. For this reason we shall only sketch the steps,
omitting the details of the proofs. In \S\ref{s4.1} we give the construction
of the master action and in \S\ref{s4.2} we discuss gauge fixing and Feynman 
rules.

\subsection{Action}  \label{s4.1}

The first step in the analysis will be to introduce new subspaces
$\oR_{g,m,n}$ of $\wt\PP_{g,m,n}$ satisfying relations similar to -- but not quite the
same -- as
\refb{eboundary}:
\ben  \label{eboundaryq}
\p \oR_{g,m,n} &=& -{1\over 2} \sum_{g_1,g_2\atop g_1+g_2=g} 
\sum_{m_1,m_2\atop m_1+m_2 = m+2}
\sum_{n_1,n_2\atop n_1+n_2 = n}
{\bf S}[\{\oR_{g_1,m_1,n_1} , \oR_{g_2,m_2,n_2}\}] \nonumber \\ &&
-{1\over 2} \sum_{g_1,g_2\atop g_1+g_2=g} 
\sum_{m_1,m_2\atop m_1+m_2 = m}
\sum_{n_1,n_2\atop n_1+n_2 = n+2}
{\bf S}[\{\oR_{g_1,m_1,n_1} ; \oR_{g_2,m_2,n_2}\}] \nonumber \\ &&
-\Delta_{NS} \oR_{g-1, m+2, n} - \Delta_R \oR_{g-1, m, n+2}
\, ,
\een
where $\Delta_{NS}$ and $\Delta_R$ are two new operations defined
as follows. $\Delta_{NS}$  takes a pair of NS punctures
on a Riemann surface corresponding to a point in $\oR_{g-1, m+2, n}$
and glues them via the special plumbing fixture relation \refb{eplumbspecial}.
$\Delta_R$ represents a similar operation on a pair of R punctures,
but we must also insert a factor of $\XX_0$ around one of the punctures.
The generalization to type II string theory is straightforward, with the
general principle that we always insert the operator $\GG$ introduced in
\refb{edefgg}, \refb{edefggii} at one of the punctures which is being glued.

Operationally the construction of $\oR_{g,m,n}$ follows a procedure 
similar to the one for $\RR_{g,m,n}$, except that now while generating 
higher genus Riemann surfaces from gluing of lower genus surfaces via
the relation \refb{eplumbing},
we also allow gluing of a pair of punctures on the same Riemann 
surface. Therefore we begin with  a three punctured sphere with arbitrary
choice of local coordinates and PCO locations consistent with 
exchange symmetries, and in the first step either glue two  
puncture on a three punctured sphere via \refb{eplumbing} 
to generate a family of
one punctured
tori, or two punctures on two three punctured spheres to generate a
family of
four punctured spheres. These generate certain subspaces of $\wt\PP_{g,m,n}$
with $(g=1,m+n=1)$ and $(g=0,m+n=4)$ whose projection to
$\MM_{g,m,n}$  generically does not
cover the whole of $\MM_{g,m,n}$. We then fill the gap with the
subspaces $\oR_{g,m,n}$ of $\wt\PP_{g,m,n}$. Again the choice of this
subspace is arbitrary except that its boundaries are fixed and it must
obey exchange and other symmetries and avoid spurious poles. 
Continuing this process we can
generate all the $\oR_{g,m,n}$'s.

Once $\oR_{g,m,n}$ 's are constructed we define new multilinear functions
$\cL A_1\cdots A_N\cR $  of $|A_1\rangle,\cdots |A_N\rangle\in \wh\HH_T$
via the relation
\be \label{edefcurlymaster}
\cL A_1\cdots A_{m+n}\cR = \sum_{g=0}^\infty (g_s)^{2g} 
\int_{\oR_{g,m,n}} \Omega^{(g,m,n)}_{6g-6+2m+2n}(|A_1\rangle,
\cdots |A_{m+n}
\rangle)\, .
\ee
We also introduce
another multilinear function 
$\sL A_2\cdots A_N\sR $ of $|A_2\rangle,\cdots |A_N\rangle\in \wh\HH_T$
taking values in $\wt\HH_T$, defined via
\be \label{edefsquaremaster}
\langle A_1| c_0^- |\sL A_2\cdots A_N\sR \rangle = \cL A_1\cdots A_N\cR 
\ee
for all $|A_1\rangle\in\wh\HH_T$.  These new functions satisfy relations 
similar to those given in \refb{ecurlysym}-\refb{eneweq2a}, except that
the right hand sides of \refb{eimpid} and \refb{emain} contain new
terms involving contraction of a pair of states inside the same bracket.
Since these relations have form identical to those given in \cite{9206084}, except
for the insertion of a $\XX_0$ operator when we contract a pair of R-sector 
states, we shall not write down these relations.

The quantum master action is given by
\be \label{emasterquantum}
S_q= g_s^{-2}\left[
-{1\over 2} \langle\wt\Psi |c_0^- Q_B \GG |\wt\Psi\rangle 
+ \langle\wt\Psi |c_0^- Q_B |\Psi\rangle + 
\sum_{n=1}^\infty {1\over n!} \cL \Psi^n\cR
\right] \, .
\ee
Following the analysis of \cite{9206084}, this can be shown to satisfy the 
quantum master equation
\be \label{eqmaster}
{1\over 2} \{S_q, S_q\} +  \Delta S_q = 0\, ,
\ee
where, for any function $F$ of the fields and anti-fields,
\be
\Delta F \equiv {\p_R\over \p\psi^s} {\p_L F\over \p \psi^*_s}\, .
\ee
The main point to note in this analysis is that on the left hand side of 
\refb{eqmaster} the $\wt\Psi$ dependent terms cancel
at the first step as in \refb{emastermani}. After this the $|\Psi\rangle$ dependent terms 
have structure identical to what appears in the closed bosonic string field theory of
\cite{9206084} except for insertion of $\XX_0$ factors on the R sector propagators.
The resulting expression can be manipulated in the same way as in 
\cite{9206084}.\footnote{There is a slightly different sign convention between 
\cite{9206084} and \cite{1408.0571,1411.7478,1501.00988}. In \cite{9206084}
the vacuum was normalized to satisfy $\langle 0| \bar c_{-1} c_{-1} \bar c_0 c_0
\bar c_1 c_1|0\rangle=1$ while in \cite{1408.0571,1411.7478,1501.00988}
 the normalization was
$\langle 0| c_{-1} \bar c_{-1} c_0 \bar c_0 
c_1 \bar c_1 e^{-2\phi}|0\rangle=1$ where $\phi$ is the bosonized 
superconformal ghost. This leads to a non-standard sign convention for
the moduli space integration measure described in \cite{1508.02481}. Alternatively
one can continue to use the standard integration measure and
include an additional factor of $(-1)^{3g-3 + N}$ in the 
definition of $\{A_1\cdots A_{N}\}$. However this difference is irrelevant for the
present analysis since the identities \refb{edefsquare}-\refb{eneweq2a} and their quantum 
generalizations  take the same form in \cite{9206084} and  
\cite{1408.0571,1411.7478,1501.00988}.}

\subsection{Gauge fixing and Feynman rules} \label{s4.2}

In the BV formalism, given the master action we 
compute the quantum amplitudes 
by carrying out the usual path integral over a Lagrangian submanifold
of the full space spanned by $\psi^r$ and $\psi^*_r$. It is most convenient
to work in the Siegel gauge
\be
b_0^+|\Psi\rangle =0, \quad b_0^+|\wt\Psi\rangle=0 \quad \Rightarrow
\quad b_0^+ \left(|\Psi\rangle
-{1\over 2} \GG|\wt\Psi\rangle\right)=0\, .
\ee
To see that this describes a Lagrangian submanifold, we divide the
basis states used in the expansion \refb{ephiexpan} into two
classes: those 
annihilated by $b_0^+$ and those annihilated by $c_0^+$. These two sets
are conjugates of each other under the inner product 
\refb{einner}. Now in the expansion given in \refb{ephiexpan}, Siegel gauge 
condition sets the coefficients of the basis states annihilated by $c_0^+$
to zero. Since in this expansion the fields and their anti-fields multiply
conjugate pairs of basis states, it follows that if the Siegel gauge condition sets
a field to zero then its conjugate anti-field remains unconstrained, and
if it sets an anti-field to zero then its conjugate field remains unconstrained.
Therefore this defines a Lagrangian submanifold.

In the Siegel gauge the propagator in $|\wt\Psi\rangle$, $|\Psi\rangle$ space
takes the form (see \cite{1508.02481} for the sign conventions)
\be 
-g_s^{2} \, b_0^+ \, b_0^- \, (L_0^+)^{-1} \,
\delta_{L_0,\bar L_0}\pmatrix{0 & 1\cr 1 & \GG}\, .
\ee
Only the lower right corner of the matrix
is important for computing amplitudes since
the interaction vertices only involve $|\Psi\rangle$ and not $|\wt\Psi\rangle$.
We can now use standard procedure  to express the different contributions
to the amplitude as integrals over subspaces of the moduli space of
punctured Riemann surfaces, and the relation  \refb{eboundaryq}
ensures that the sum over all Feynman diagrams cover the whole moduli
space\cite{GMW,9206084}. 
Note that only states in $\wh\HH_T$ propagate along internal lines
but they can carry arbitrary ghost number.\footnote{Some qualification
is warranted here. The states which enter the vertex are states in $\wh\HH_T$
annihilated by $b_0^+$. But the propagator itself consists of the 
operator $-g_s^{2}  \, b_0^+ b_0^- (L_0^+)^{-1} \GG$ sandwiched between
a pair of basis states in the conjugate
sector, which are states in $\HH$ carrying picture numbers $(-1,-3/2)$
and annihilated by $c_0^+$ and $c_0^-$. Since there is no restriction on
the ghost number, there are apparently infinite number of states at each
mass level obtained by repeated application of the zero mode $\beta_0$ of
$\beta$ in the R sector. However the operator $\XX_0$ in the propagator
annihilates all but a finite number of these states. This can be seen using the fact that
the application of $\beta_0$ reduces the ghost number of the state.
On the other hand the application of $\XX_0$
produces a state of picture number $-1/2$ for which $\beta_0$ annihilates the
vacuum and hence at a given level, we can no longer 
have states of arbitrarily small (i.e.\ large negative)
ghost number.
Therefore a state of picture number $-3/2$ must be annihilated by $\XX_0$ for
sufficiently small ghost number since there will be no candidate state with
the right quantum numbers.  This in turn shows that only a finite
number of states propagate at each mass level. This property is manifest if instead
of $\XX_0$ we use the kinetic operator used in \cite{1508.00366}, but at this stage
it is not clear how to write down a fully gauge invariant closed string field theory action
based on this kinetic operator.}

Since the field $|\wt\Psi\rangle$ continues to appear only in the kinetic term
even in the full quantum BV action, 
the additional modes we
have introduced via $|\wt\Psi\rangle$ decouple from the interacting 
part of the theory.
Indeed
1PI amplitudes computed using the master action would reproduce the
1PI action given in \refb{e1PIact}, with the only difference that
the $\RR_{g,m,n}$'s
involved in the definitions of $\{A_1\cdots A_N\}$ are not defined
independently, but constructed from the $\oR_{g,m,n}$'s used for
defining $\cL A_1\cdots A_N\cR$ by plumbing fixture of $\oR_{g,m,n}$'s
in all possible ways via the relation \refb{eplumbing}, 
but keeping only the `1PI contributions'.
In \refb{e1PIact} 
$|\wt\Psi\rangle$ appears only in the kinetic term, showing that
its equation of motion leads to free fields even in the full quantum theory.

\bigskip

\noindent {\bf Acknowledgement:}
We wish to thank Nathan Berkovits, Ted Erler, Yuji Okawa, Martin Schnabl 
and Barton Zwiebach for useful discussions.
This work  was
supported in part by the 
DAE project 12-R\&D-HRI-5.02-0303 and J. C. Bose fellowship of 
the Department of Science and Technology, India.




\begin{thebibliography}{99}

\bibitem{1209.5461} 
  E.~Witten,
  ``Superstring Perturbation Theory Revisited,''
  arXiv:1209.5461 [hep-th].


\bibitem{1304.2832}
E.~Witten,
``More On Superstring Perturbation Theory,''
  arXiv:1304.2832 [hep-th].

\bibitem{1311.1257} 
  R.~Pius, A.~Rudra and A.~Sen,
  ``Mass Renormalization in String Theory: Special States,''
  arXiv:1311.1257 [hep-th].

\bibitem{1401.7014} 
  R.~Pius, A.~Rudra and A.~Sen,
  ``Mass Renormalization in String Theory: General States,''
  arXiv:1401.7014 [hep-th].

\bibitem{1404.6254} 
  R.~Pius, A.~Rudra and A.~Sen,
  ``String Perturbation Theory Around Dynamically Shifted Vacuum,''
  arXiv:1404.6254 [hep-th].

\bibitem{1408.0571} 
  A.~Sen,
  ``Off-shell Amplitudes in Superstring Theory,''
  arXiv:1408.0571v4 [hep-th].
  
\bibitem{1411.7478} 
  A.~Sen,
  ``Gauge Invariant 1PI Effective Action for Superstring Field Theory,''
  arXiv:1411.7478 [hep-th].




\bibitem{1501.00988} 
  A.~Sen,
  ``Gauge Invariant 1PI Effective Superstring Field Theory: Inclusion of the Ramond Sector,''
  arXiv:1501.00988 [hep-th].

\bibitem{1508.02481} 
  A.~Sen,
  ``Supersymmetry Restoration in Superstring Perturbation Theory,''
  arXiv:1508.02481 [hep-th].

\bibitem{wittensft} 
  E.~Witten,
  ``Noncommutative Geometry and String Field Theory,''
  Nucl.\ Phys.\ B {\bf 268}, 253 (1986).

\bibitem{thorn} 
  C.~B.~Thorn,
  ``String Field Theory,''
  Phys.\ Rept.\  {\bf 175}, 1 (1989).


\bibitem{bochicchio} 
  M.~Bochicchio,
  ``Gauge Fixing for the Field Theory of the Bosonic String,''
  Phys.\ Lett.\ B {\bf 193}, 31 (1987).


\bibitem{saadi} 
  M.~Saadi and B.~Zwiebach,
  ``Closed String Field Theory from Polyhedra,''
  Annals Phys.\  {\bf 192}, 213 (1989).

 \bibitem{kugo}
T.~Kugo, H.~Kunitomo and K.~Suehiro,
``Nonpolynomial Closed String Field Theory,''
  Phys.\ Lett.\ B {\bf 226}, 48 (1989);
T.~Kugo and K.~Suehiro,
``Nonpolynomial Closed String Field Theory: Action and Its Gauge Invariance,''
  Nucl.\ Phys.\ B {\bf 337}, 434 (1990).


\bibitem{9206084} 
  B.~Zwiebach,
  ``Closed string field theory: Quantum action and the B-V master equation,''
  Nucl.\ Phys.\ B {\bf 390}, 33 (1993)
  [hep-th/9206084].

\bibitem{9301097} 
  H.~Hata and B.~Zwiebach,
  ``Developing the covariant Batalin-Vilkovisky approach to string theory,''
  Annals Phys.\  {\bf 229}, 177 (1994)
  [hep-th/9301097].

\bibitem{wittenssft} 
  E.~Witten,
  ``Interacting Field Theory of Open Superstrings,''
  Nucl.\ Phys.\ B {\bf 276}, 291 (1986).
  

  \bibitem{9503099}
  N.~Berkovits,
  ``SuperPoincare invariant superstring field theory,''
  Nucl.\ Phys.\ B {\bf 450} (1995) 90
   [Erratum-ibid.\ B {\bf 459} (1996) 439]
  [hep-th/9503099].

\bibitem{0109100}
  N.~Berkovits,
  ``The Ramond sector of open superstring field theory,''
  JHEP {\bf 0111} (2001) 047
  [hep-th/0109100].

\bibitem{0406212}
  Y.~Okawa and B.~Zwiebach,
  ``Heterotic string field theory,''
  JHEP {\bf 0407} (2004) 042
  [hep-th/0406212].

\bibitem{0409018}
  N.~Berkovits, Y.~Okawa and B.~Zwiebach,
  ``WZW-like action for heterotic string field theory,''
  JHEP {\bf 0411} (2004) 038
  [hep-th/0409018].

\bibitem{1201.1761} 
  M.~Kroyter, Y.~Okawa, M.~Schnabl, S.~Torii and B.~Zwiebach,
  ``Open superstring field theory I: gauge fixing, ghost structure, and propagator,''
  JHEP {\bf 1203}, 030 (2012)
  [arXiv:1201.1761 [hep-th]].

\bibitem{1305.3893} 
  H.~Matsunaga,
  ``Construction of a Gauge-Invariant Action for Type II Superstring Field Theory,''
  arXiv:1305.3893 [hep-th].

\bibitem{1312.2948}
  T.~Erler, S.~Konopka and I.~Sachs,
  ``Resolving Witten`s superstring field theory,''
  JHEP {\bf 1404} (2014) 150
  [arXiv:1312.2948 [hep-th]].

\bibitem{1312.7197}
  H.~Kunitomo,
  ``The Ramond Sector of Heterotic String Field Theory,''
  PTEP {\bf 2014} 4,  043B01
  [arXiv:1312.7197 [hep-th]].

 \bibitem{1403.0940}
  T.~Erler, S.~Konopka and I.~Sachs,
 ``NS-NS Sector of Closed Superstring Field Theory,''
  arXiv:1403.0940 [hep-th].

\bibitem{1407.8485} 
  H.~Matsunaga,
  ``Nonlinear gauge invariance and WZW-like action for NS-NS superstring field theory,''
  arXiv:1407.8485 [hep-th].
  
\bibitem{1412.5281} 
  H.~Kunitomo,
  ``Symmetries and Feynman Rules for Ramond Sector in Open Superstring Field Theory,''
  arXiv:1412.5281 [hep-th].
  
 \bibitem{1505.01659} 
  T.~Erler, Y.~Okawa and T.~Takezaki,
  ``$A_\infty$ structure from the Berkovits formulation of open superstring field theory,''
  arXiv:1505.01659 [hep-th].
  
\bibitem{1505.02069} 
  T.~Erler,
  ``Relating Berkovits and $A_\infty$ Superstring Field Theories; Small Hilbert Space Perspective,''
  arXiv:1505.02069 [hep-th].

\bibitem{1506.05774} 
  T.~Erler, S.~Konopka and I.~Sachs,
  ``Ramond Equations of Motion in Superstring Field Theory,''
  arXiv:1506.05774 [hep-th].

\bibitem{1506.06657} 
  K.~Goto and H.~Matsunaga,
  ``On-shell equivalence of two formulations for superstring field theory,''
  arXiv:1506.06657 [hep-th].

\bibitem{1508.00366} 
  H.~Kunitomo and Y.~Okawa,
  ``Complete action of open superstring field theory,''
  arXiv:1508.00366 [hep-th].

\bibitem{FMS} 
  D.~Friedan, E.~J.~Martinec and S.~H.~Shenker,
  ``Conformal Invariance, Supersymmetry and String Theory,''
  Nucl.\ Phys.\ B {\bf 271}, 93 (1986).

 \bibitem{9711087} 
  N.~Berkovits and B.~Zwiebach,
  ``On the picture dependence of Ramond-Ramond cohomology,''
  Nucl.\ Phys.\ B {\bf 523}, 311 (1998)
  [hep-th/9711087].
  
  \bibitem{1504.00609} 
  A.~Sen and E.~Witten,
  ``Filling The Gaps With PCO's,''
  arXiv:1504.00609 [hep-th].

\bibitem{0412215} 
  Y.~Michishita,
  ``A Covariant action with a constraint and Feynman rules for fermions in open superstring field theory,''
  JHEP {\bf 0501}, 012 (2005)
  [hep-th/0412215].

\bibitem{Verlinde:1987sd} 
  E.~P.~Verlinde and H.~L.~Verlinde,
  ``Multiloop Calculations in Covariant Superstring Theory,''
  Phys.\ Lett.\ B {\bf 192}, 95 (1987).

 \bibitem{lechtenfeld}
O.~Lechtenfeld,
``Superconformal Ghost Correlations On Riemann Surfaces,''
  Phys.\ Lett.\ B {\bf 232}, 193 (1989).

\bibitem{morozov}
  A.~Morozov,
  ``STRAIGHTFORWARD PROOF OF LECHTENFELD'S FORMULA FOR BETA, gamma CORRELATOR,''
  Phys.\ Lett.\ B {\bf 234}, 15 (1990)
  [Yad.\ Fiz.\  {\bf 51}, 301 (1990)]
  [Sov.\ J.\ Nucl.\ Phys.\  {\bf 51}, 190 (1990)].
  
\bibitem{GMW} 
  S.~B.~Giddings, E.~J.~Martinec and E.~Witten,
  ``Modular Invariance in String Field Theory,''
  Phys.\ Lett.\ B {\bf 176}, 362 (1986).

\end{thebibliography}
\end{document}